# A phenomenology condition other than zero resistance and a possible pairing mechanism of holes-electrons for superconductivity

Wei-long She *

*School of Physics, Sun Yat-Sen University, Guangzhou 510275, China*

The underlying mechanism of unconventional high-temperature superconductivity is a great challenge to condensed matter physics. However, zero dissipation of electric current is the commonness of superconductors whether they are conventional or unconventional ones. In this presentation, the Ohm law in a nonmagnetic conductor is derived from a set of modified electromagnetic equations that involve Maxwell ones. It is found that, the steady current dissipation in a conductor can be expressed as $\mathbf{J}\cdot\mathbf{E}=c^2\rho/(u\varepsilon_r\mu_r)$, where **J**, **E**, $\rho$, c, u, $\varepsilon_r$ and $\mu_r$ are the electric current density, electric field strength, free electric charge density, light speed in vacuum, effective mobility of carriers, relative dielectric constant and permeability, respectively. This relation indicates that, in a steady state of $\mathbf{J}\neq 0$, if $\rho=0$ then $\mathbf{J}\cdot\mathbf{E}=0$ and the conductor comes into a superconducting state. It is also found that the condition $\rho=0$ is valid for superconductivity of magnetic materials and is a sufficient than necessary one. When $\rho=0$ the $\mathbf{E}$ (involving the Hall electric field strength) becomes zero, which solves the pending problem why vanishing of Hall-effect in some superconducting states, besides, suggests a superconductive pairing mechanism of holes and electrons. Two examples of superconducting state under the condition $\rho=0$ are discussed.

* Email: shewl@mail.sysu.edu.cn

## 1. **Introduction**

Almost four decades ago, the first family of unconventional high-$T_c$ superconductors, the Cu-based ones (cuprates) [1], was discovered. Since then, enormous efforts have been devoted to find high-$T_c$ superconductive materials. Besides the cuprates [2], the iron-based [3-5] and nickel-based [6-8] ones have been found one after the other. However, the understanding of the superconductivity with these unconventional high-temperature superconductors is very difficult. Although there are Bardeen-Cooper-Schrieffer [9] and Migdal-Éliashberg [10,11] theories, extended s-wave [12,13] and d-wave pairing theories [14,15], Hubbard [16-18] and t-J models [19,20], variational theory [21]，and mean-field theory of strongly correlated fermion systems [22], the underlying mechanisms of these unconventional high-temperature superconductors still remain unknown and the microscopic theory of high-temperature superconductivity is yet to be established [8,23-25]. To characterize high-$T_c$ superconductors, avoiding the microcosmic mechanisms, here we explore a phenomenology condition other than zero resistance for superconductivity.

It is known that zero dissipation of electric current is the commonness of superconductors. And the dissipation of current in a conductor can be described by $\mathbf{J}\cdot\mathbf{E}$ . In this paper, a set of modified electromagnetic equations involving Maxwell ones are proposed. From the equations the Ohm law (which is independent of Maxwell equations in classical electromagnetics) in a nonmagnetic conductor is derived; and it is found that, the dissipation of current can be expressed as $\mathbf{J}\cdot\mathbf{E}=c^2\rho/(u\varepsilon_r\mu_r)$ . This indicates that, in a steady state of $\mathbf{J}\neq 0$ , as $\rho=0$ ， the conductor becomes non-dissipative and comes into a superconducting state. It is also found that the condition is valid for magnetic materials. Two examples of superconducting state under the condition will be shown. But, the condition $\rho = 0$ is a sufficient than necessary one for superconductivity. For example, the superconducting state described by London theory [26] can be with $\rho\neq 0$ but $\mathbf{J}\cdot\mathbf{E}=0$ .

The experimental observations had found that the Hall coefficients of some materials will vanish when they go into superconducting states [27-30]. Lewis concluded that "this property lies outside the existing body of theory, and the general requirements it imposes on a future theory are adduced" [31]. We will show that the pending problem can be solved under the condition $\rho=0$ , besides; the condition suggests a superconductive pairing mechanism of holes and electrons.

## **2. Modified electromagnetic equations**

The modified electromagnetic equations proposed are in the forms (see its derivation in Appendix)

$$\nabla \times \mathbf{E} = -\frac{\partial \mathbf{B}}{\partial t}, \qquad (1)$$

$$\nabla \cdot \mathbf{B} = 0 \ , \qquad (2)$$

$$\nabla \times \mathbf{H} = \mathbf{J} + \frac{\partial \mathbf{D}}{\partial t}, \qquad (3)$$

$$\nabla \cdot \mathbf{D} = \rho \ , \qquad (4)$$

$$\rho = \frac{k}{c^2 \sqrt{\mu_r \varepsilon_r}} (\varphi - \frac{\partial \phi}{\partial t}) \ , (5)$$

$$\mathbf{J} = \frac{k}{\mu_r \varepsilon_r \sqrt{\mu_r \varepsilon_r}} (\mathbf{A} + \nabla \phi) \ , (6)$$

Where **E**, **D**, **B**, and **H** are the electric field strength, electric displacement, magnetic induction and magnetic field strength, respectively; $\mathbf{A}$ and $\varphi$ are the vector potential and scalar potential, respectively; $\phi$ is a scalar function; $k$ is a parameter depending on space coordinates and time. Equations (1)-(4) are Maxwell's and (5)-(6) are new ones. And Eq.(6) is consistent with the Ginzburg and Landau equations [32]. If $k = 0$, then $\mathbf{J} = 0$ and $\rho = 0$. What we should consider are only Maxwell Equations (1)-(4). Assuming $k \neq 0$ and using both relations $\mathbf{B} = \nabla \times \mathbf{A}$ and $\mathbf{E} = -\partial \mathbf{A} / \partial t - \nabla \varphi$, Eqs. (5) and (6) yield

$$\frac{\partial (\alpha \mathbf{J})}{\partial t} + \nabla (\frac{c^2 \alpha \rho}{\mu_r \varepsilon_r}) = -\mathbf{E} \ , \qquad (7)$$

$$\nabla \times (\alpha \mathbf{J}) = \mathbf{B} \ , \qquad (8)$$

where $\alpha = (\mu_r \varepsilon_r \sqrt{\mu_r \varepsilon_r}) / k$. One can see that Eqs. (7) and (8) involve Eqs. (1) and (2). We now consider a steady state and assume $k \neq 0$ but $\rho = 0$, then we have

$$\mathbf{E} = -\nabla (\frac{c^2 \alpha \rho}{\mu_r \varepsilon_r}) = 0 \qquad (9)$$

Equation (9) means zero dissipation of current since $\mathbf{J} \cdot \mathbf{E} = 0$. In this case, the conductor comes into a superconducting state. It should be noticed that condition $\rho = 0$ is a sufficient than necessary one for superconductivity. To show this clearly, we will derive the Ohm law from modified electromagnetic equations and reveal the relation between dissipation and free electric charge.

### 3. Deduction of Ohm's law and condition $\rho = 0$ (or $\alpha = \text{constant}$) for superconductivity

In the classical electromagnetics, the Ohm law is independent of Maxwell equations. However, we will show that the Ohm law can be derived from the above modified electromagnetic equations.

Let's consider an infinitely long straight circular wire carrying a steady current **J**. For convenience, we suppose that the wire is isotropic and nonmagnetic with $\mu_r = 1$ . In this case, the modified electromagnetic equations become

$$\nabla\left(\frac{\alpha\rho}{\varepsilon_r}\right) = -\frac{\mathbf{E}}{c^2}, \qquad (10)$$

$$\nabla \times (\alpha \mathbf{J}) = \mathbf{B}, \qquad (11)$$

$$\nabla \times \mathbf{H} = \mathbf{J}, \qquad (12)$$

$$\nabla \cdot (\varepsilon_0 \varepsilon_r \mathbf{E}) = \rho, \qquad (13)$$

where $\varepsilon_0$ is dielectric constant in vacuum. We choose a circular cylindrical coordinates with z-axis along the direction of the current **J.** Obviously**, J** has only a component $J_z$. Denoting it as J and using $\nabla \cdot \mathbf{J} = 0$ , we have

$$\frac{\partial J}{\partial z} = 0, \qquad (14)$$

which means that J is independent of *z.* From Eqs. (11) and (12), we can deduce

$$\nabla(\nabla \cdot (\alpha \mathbf{J})) - \nabla^2 (\alpha \mathbf{J}) = \mu_0 \mathbf{J}, \ (15)$$

where $\mu_0$ is the permeability in vacuum. Its expansion is

$$\frac{\partial}{\partial r}\frac{\partial}{\partial z}(\alpha J)\mathbf{e}_r + \frac{1}{r}\frac{\partial^2}{\partial \theta \partial z}(\alpha J)\mathbf{e}_\theta - \left[\frac{1}{r}\frac{\partial}{\partial r}\left(r\frac{\partial(\alpha J)}{\partial r}\right) + \frac{1}{r^2}\frac{\partial^2}{\partial \theta^2}\right]\mathbf{e}_z = \mu_0 J \mathbf{e}_z, \ (15a)$$

where $\mathbf{e}_r$ , $\mathbf{e}_\theta$ and $\mathbf{e}_z$ are three unit-vectors of circular cylindrical coordinates, respectively. This vector equation involves three scalar ones:

$$\frac{\partial}{\partial r}\frac{\partial}{\partial z}(\alpha J) = 0, \qquad (16)$$

$$\frac{1}{r}\frac{\partial^2}{\partial \theta \partial z}(\alpha J) = 0, \qquad (17)$$

$$-\frac{1}{r}\frac{\partial}{\partial r}\left(r\frac{\partial(\alpha J)}{\partial r}\right) - \frac{1}{r^2}\frac{\partial^2}{\partial \theta^2}(\alpha J) = \mu_0 J, \ (18)$$

Equations (16) and (17) give

$$J\left(\frac{\partial \alpha}{\partial z}\right) = y(z), \qquad (19)$$

where $y(z)$ is an arbitrary function depending on *z.* There are two cases: $y(z) = 0$ as $\partial\alpha / \partial z = 0$ and $y(z) \neq 0$ as $\partial\alpha / \partial z \neq 0$ . Let us recall Eq. (10), which includes these three scalar equations

$$\frac{\partial}{\partial r}\left(\frac{\alpha\rho}{\varepsilon_r}\right) = -\frac{E_r}{c^2}, \qquad (20)$$

$$\frac{\partial}{\partial z}\left(\frac{\alpha\rho}{\varepsilon_r}\right) = -\frac{E_z}{c^2}, \qquad (21)$$

$$\frac{1}{r}\frac{\partial}{\partial\theta}(\frac{\alpha\rho}{\varepsilon_r})=-\frac{E_\theta}{c^2}\ . \qquad (22)$$

Supposing $E_z \neq 0$ then $\rho \neq 0$ and $\partial\alpha/\partial z \neq 0$ (see below). It is helpful to consider the symmetry of the system before going forwards. The symmetry of system requires $J$ , $E_r$ and $E_z$ to be independent of $\theta$ and *z*. On the other hand, by (10) we have

$$\nabla\times\mathbf{E}=0\rightarrow(\frac{1}{r}\frac{\partial E_z}{\partial\theta}-\frac{\partial E_\theta}{\partial z})\mathbf{e}_r+(\frac{\partial E_r}{\partial z}-\frac{\partial E_z}{\partial r})\mathbf{e}_\theta+(\frac{1}{r}\frac{\partial(rE_\theta)}{\partial r}-\frac{1}{r}\frac{\partial E_r}{\partial\theta})\mathbf{e}_z=0\ . \qquad (23)$$

This leads to that $E_\theta=0$ and $E_z$ is independent of r. Further, by Eq.(13), $\rho$ is independent of $\theta$ and *z*. By Eq. (22), $(\alpha\rho)$ and $\alpha$ are independent of $\theta$ . With these conditions Eq. (21) becomes

$$\frac{\alpha\rho}{\varepsilon_r}=-\frac{E_z}{c^2}z+g(r) \quad \text{or} \quad \alpha=-\frac{\varepsilon_r E_z}{c^2\rho}z+h(r)\ , \qquad (24)$$

where *g*(r) and *h*(r) are two arbitrary functions depending on *r*. Then we have

$$\frac{\partial\alpha}{\partial z}=-\frac{\varepsilon_r E_z}{c^2\rho} \qquad (25)$$

Using Eqs. (19) and (24) we obtain

$$J=-\frac{Dc^2\rho}{\varepsilon_r E_z} \qquad (26)$$

Here one would find that $D[=y(z)]$ is a constant since *J,* $E_z$ , $\rho$ and $\varepsilon_r$ are all independent of *z*. Besides, $D$ cannot be zero (i.e., $\partial\alpha/\partial z\neq 0$ ) or else $J=0$ . Equation (26) can be rewritten as

$$\rho=-\frac{J\varepsilon_r E_z}{Dc^2}\ . \qquad (27)$$

Since $J$ is independent of $\theta$ we have $\partial(\alpha J)/\partial\theta=0$ . Substituting Eq. (27) into Eq. (18) we obtain

$$\frac{1}{r}\frac{\partial}{\partial r}(r\frac{\partial}{\partial r}(\alpha\frac{Dc^2\rho}{\varepsilon_r E_z}))=\mu_0\frac{Dc^2\rho}{\varepsilon_r E_z}\ . \qquad (28)$$

Since D and $E_z$ are independent of r, Eq. (28) becomes

$$\frac{1}{r}\frac{\partial}{\partial r}(r\frac{\partial}{\partial r}(\frac{\alpha\rho}{\varepsilon_r}))=-\mu_0\frac{\rho}{\varepsilon_r}\ . \qquad (29)$$

Substituting Eq. (20) into (29) we get

$$\frac{1}{r}\frac{\partial}{\partial r}(rE_r)=\mu_0\frac{c^2\rho}{\varepsilon_r}\ . \qquad (30)$$

By using Eq. (27), Eq. (30) can be rewritten as

$$\frac{1}{r}\frac{\partial}{\partial r}(rE_r)=-\frac{\mu_0 E_z}{D}J\rightarrow\frac{1}{r}\frac{\partial}{\partial r}(-rE_r\frac{D}{E_z})=\mu_0 J\ . \qquad (31)$$

And by Eq. (12) we can get

$$\frac{1}{r}\frac{\partial}{\partial r}(rB_\theta) = \mu_0 J \,. \qquad (32)$$

Here we have used the relation of $\mathbf{B} = B_\theta \mathbf{e}_\theta$. Comparing Eq. (31) with Eq. (32), we find

$$E_r = -\frac{E_z}{D}B_\theta \,. \qquad (33)$$

On the other hand, zero current along the direction of $\mathbf{e}_r$ requires

$$eE_r = evB_\theta \to neE_r = JB_\theta\,(J = nev) \;, \qquad (34)$$

where $e$, $v$ and $n$ are the charge, average speed and effective density of carrier, respectively. Comparing Eq. (33) with Eq. (34), we reach

$$J = -\frac{(ne)E_z}{D} = (neu)E_z \qquad (35)$$

This is the right Ohm law, where $u = -1/D$ is effective mobility of carrier. With Eqs.(27) and (35) the free electric charge density can be expressed as $\rho = \varepsilon_r J^2/(nec^2)$ ,which tells us a small number density of free electric charge. For example, for a copper wire (with carrier density $n_c = 8.4\times10^{28}/m^3$ ) carrying a steady current $J = 2.4\times10^7 A/m^2$ , the number density of free electric charge calculated to be $3.0\times10^7/m^3 (<< n_c)$ , where $\varepsilon_r = 10$ is assumed**.** By Eq.(26) we obtain the Joule heat density

$$Q = JE_z = \frac{c^2\rho}{u\varepsilon_r} \;\text{ or }\; Q = \frac{c^2\rho}{u\varepsilon_r\mu_r}\;(\text{as } \mu_r \neq 1) \quad (36)$$

The equation shows $Q \propto 1/u$ , which is consistent with the current view of dissipation. However, it also indicates that Joule heat density depends on the free electric charge density. One sees that as $\rho \to 0$ the Joule heat will vanish [ $J\cdot 0 = -c^2 D/(\varepsilon_r\mu_r)\cdot 0 = 0$ according to Eqs.(19) and (21)]. In this case, the conductor comes into a superconducting state, which suggests an alternative window for prying into the mechanisms of unconventional high-temperature superconductors. The condition is also valid for the superconductivity of magnetic materials since as $\rho = 0$ , Eq.(9) gives $\mathbf{E} = 0$ then $\mathbf{J}\cdot\mathbf{E} = 0$ for both magnetic and nonmagnetic materials.

Note that $E_r$ is the electric field strength of self-Hall-effect. So Eq. (34) can be rewritten as $E_r = R_H J B_\theta$ , where $R_H$ is the Hall coefficient. If $R_H = 0$ , then $E_r = 0$ and $\rho = 0$ [see Eq.(30)]. It is true the other way round, i.e., when $\mathbf{J} \neq 0$ and $\rho = 0$ , by Eq.(9) it must be $\mathbf{E} = 0$ ,then all electric fields vanish. So, $E_r = 0$ and $R_H = 0$ , which appears to be consistent with the experiments [27-30]. This would solve the pending theoretical problem [31] why vanishing of Hall-effect in some superconducting states.

It is known that when the carriers involve both electrons and holes, under the condition of weak magnetic field the Hall coefficient can be expressed as

$$R_H = \frac{pu_h^2 - nu_e^2}{q(pu_h + nu_e)^2}, \qquad (37)$$

where $p, u_h$ ( $n, u_e$ ) are the density and mobility of holes (electrons), respectively; and $q$ is the basic charge. The experimental observations have revealed that the Hall coefficients of some high-temperature superconductors will change their signs near entering the superconducting states [28-30], which suggest that there exists the transmission competition between holes and electrons in superconductors. One sees that if create a condition, which makes $pu_h^2 - nu_e^2 = 0$ , then $R_H = 0$ and $\rho = 0$ , the conductors will come into a superconducting state. This suggests a superconductive pairing mechanism of holes and electrons. It should be noticed that as $\rho = 0$ , Eq (21) gives $E_z = 0$ , and Eq.(25) becomes an uncertain one $(\partial / \partial z)\alpha = 0/0$ . We cannot deduce Eqs.(26) and (33), indeed not Ohm law Eq.(35). This suggests that $\rho \to 0$ is a process of phase transition.

We emphasize here that the condition $\rho = 0$ is a sufficient than necessary one for superconductivity. For example, even if $\rho \neq 0$ , if $\alpha = \text{constant} = -m/(q^2 n^*)$ [$m$ and $n^*$ are the effective mass and density of carrier(s), respectively] , i.e., the superconducting state is described by London theory [26], Eq. (18) becomes as

$$-\frac{1}{r}\frac{\partial}{\partial r}(r\frac{\partial J}{\partial r}) = \frac{\mu_0}{\alpha} J , \qquad (38)$$

From which one can obtain immediately the solution $J = J_0 I_0(r/\lambda)/I_0(R/\lambda)$ , where $R$ is the radius of the wire, $I_0(x)$ is the 0-order modified Bessel function of the first kind; $\lambda = \sqrt{-\alpha/\mu_0}$ , the penetration depth; and $J_0$ is the electric current density at $r$ = R. Under the condition of $\alpha = \text{constant}$ and Eqs.(10) and (13), it is easy to deduce that $\rho = \rho_0 I_0(r/\lambda)/I_0(R/\lambda)$ , where $\rho_0$ is the free electric charge density at $r$ = R. Further, Eqs.(20)-(22) gives $\mathbf{E} = E_r \mathbf{e}_r$ , which leads to $\mathbf{J} \cdot \mathbf{E} = (J_z \mathbf{e}_z) \cdot (E_r \mathbf{e}_r) = 0$ . The Joule heat vanishes. So the condition ρ = 0 is a sufficient than necessary one for superconductivity. It should be noticed that in the case of $\alpha = \text{constant}$ , Eq.(19) becomes $J(\partial \alpha / \partial z) = 0$ . We cannot deduce Eq.(26) and the Ohm law Eq.(35) any more. Besides, $\rho \neq 0$ as long as $E_r \neq 0$ ( $R_H \neq 0$ ), which seems that the conditions $\alpha = \text{constant}$ and $\rho = 0$ depict respectively two different superconducting states. However, the line lying between presence and absence of Ohm law is still the boundary of normal conductors and superconductors.

**4. Two examples of superconducting state under condition $\rho = 0$**

In the steady state, when $\mathbf{J} \neq 0$ but $\rho = 0$ ( $\mathbf{E} = 0$ ) the electromagnetic equations leave two forms:

$$\nabla \times \mathbf{H} = \mathbf{J}, \qquad (39)$$

$$\nabla \times (\alpha \mathbf{J}) = \mathbf{B}. \qquad (40)$$

With Eqs.(39) and (40) we now discuss two examples of superconducting state. Below, the conductors are supposed to be isotropic and nonmagnetic with $\mu_r = 1$.

Example 1：By considering an infinitely long straight circular wire of radius $R$ and with a steady current **J,** choosing a circular cylindrical coordinates with z-axis along the direction of the current **J,** Eq.(18) can be derived from Eqs.(39) and (40). Note that as $\rho = 0$ , Eq (21) gives $E_z = 0$ , and Eqs.(24) and (25) are invalid, which suggests to be $\partial\alpha / \partial z = 0$ , i.e., $\alpha$ is independent of $z$ . But, Eqs. (16) and (17) allow $\alpha$ to be as a function of $r$ , further, as a function of $J$. We now take $\alpha$ as a function of $J$. Then Eq.(18) becomes

$$-\frac{1}{r}\frac{\partial}{\partial r}\left(r\frac{\partial(\alpha J)}{\partial r}\right) = \mu_0 J \quad (41)$$

We may consider the simplest case, i.e., $\alpha = \alpha_0 + \alpha_1 J$ , where $\alpha_0$ and $\alpha_1$ are two constants, respectively. By letting $\lambda = \sqrt{-\alpha_0 / \mu_0}$ , $\beta = \alpha_1 J_0 / (\alpha_0 I_0(R/\lambda))$ , $j = J \cdot I_0(R/\lambda) / J_0$ ( $J_0$ is the electric current density at $r$=R) and $r' = r/\lambda$ (still signed by r), Eq.(41) reduces to

$$r\frac{d^2 j}{dr^2} + \beta r \frac{d^2 j^2}{dr^2} + \beta \frac{dj^2}{dr} + \frac{dj}{dr} - rj = 0 \quad (42)$$

Equation (42) can be solved by approximation method: First, expand $j$ as a series of $r$ , i.e.,

$$j(r) = a_0 + a_1 r + a_2 r^2 + a_3 r^3 + \ldots \quad . \qquad (43)$$

Then, substitute Eq.(43) into Eq.(42) and use the optimization method to determine the coefficients in Eq.(43). The results for several values of $\beta$ are shown in Fig. 1. One sees that the superconducting current tends to the surface of the wire. For a positive $\beta$ , the varying of current is slower than that described by London theory; and for a negative $\beta$ , the varying is faster than that.

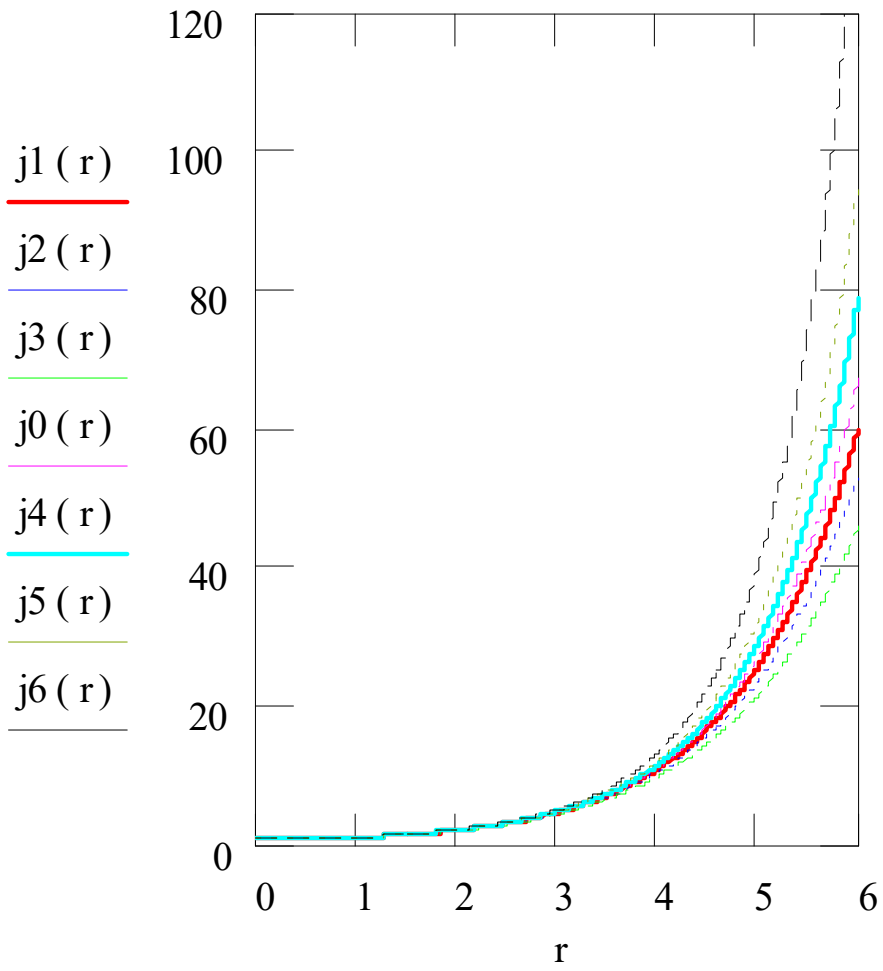

Fig.1 The numerical results of superconducting currents for different values of parameter $\beta$, where currents $j0(r), j1(r), j2(r), j3(r), j4(r), j5(r)$ and $j6(r)$ correspond to $\beta$ with values of $0,\ 0.001,\ 0.005,\ 0.01,\ -0.002,\ -0.005$ and $-0.01$, respectively. And $j0(r) = I_0(r)$, the 0-order modified Bessel function of the first kind.

Example 2：Assuming that the external magnetic field $\mathbf{H}_0$ is directed along the Z axis and $\mu_r = 1$, considering the situation of 2-dimensions, by Eqs. (39) and (40) we can deduce

$$\frac{\partial^2 B}{\partial X^2} + \frac{\partial^2 B}{\partial Y^2} + \frac{\mu_0}{\alpha} B + \frac{1}{\alpha}\left(\frac{\partial \alpha}{\partial X}\frac{\partial B}{\partial X} + \frac{\partial \alpha}{\partial Y}\frac{\partial B}{\partial Y}\right) = 0\,, \qquad (44)$$

where $B = |\mathbf{B}|$. Let $\alpha = 1/\beta$, Eq. (44) becomes

$$\frac{\partial^2 B}{\partial X^2} + \frac{\partial^2 B}{\partial Y^2} + \mu_0 \beta B - \frac{1}{\beta}\left(\frac{\partial \beta}{\partial X}\frac{\partial B}{\partial X} + \frac{\partial \beta}{\partial Y}\frac{\partial B}{\partial Y}\right) = 0\,. \qquad (45)$$

By introducing a parameter $\lambda_0^2 = m/(\mu_0 q^2 n^*)$, the equation is reduced to a dimensionless one

$$\frac{\partial^2 h}{\partial x^2} + \frac{\partial^2 h}{\partial y^2} - fh - \frac{1}{f}\left(\frac{\partial f}{\partial x}\frac{\partial h}{\partial x} + \frac{\partial f}{\partial y}\frac{\partial h}{\partial y}\right) = 0\,, \qquad (46)$$

where $f = -\mu_0 \lambda_0^2 \beta$ and $h = B/(\mu_0 \sqrt{2} H_c)$ ($h$ is reduced magnetic field and $H_c$ is the critical magnetic field); $x = X/\lambda_0$ and $y = Y/\lambda_0$.

We now show that Eq.(46) describes the magnetic field at the mixed state of type II superconductors. Suppose that $h$ and $f$ is connected by such a linear relation

$$h = h_a - \frac{f}{2\kappa}\,, \qquad (47)$$

where $h_a$ is the reduced external magnetic field and $\kappa = 2\pi/\phi_0$ ($\phi_0$ is a fluxon). With this relation Eq. (46) can be rewritten as

$$\frac{\partial^2 \ln f}{\partial x^2}+\frac{\partial^2 \ln f}{\partial y^2}+2\kappa h_a - f = 0 \quad . \qquad (48)$$

According to Eq. (47), if $h$ is very close to $h_a$ and $h_a$ is very close to reduced upper critical field of type II superconductors, making $2\kappa h_a >> f$ so that the term $-f$ in Eq. (48) can be ignored, we have

$$\frac{\partial^2 \ln f}{\partial x^2}+\frac{\partial^2 \ln f}{\partial y^2}+2\kappa h_a = 0 \, . \quad (49)$$

It is easy to verify that this function

$$f(x,y)=\left|C_0 e^{-\frac{\kappa h_a}{2}y^2}\sum_{n=-\infty}^{\infty}\exp[i\frac{n(n-1)\pi b\cos\theta}{a}]\exp[\frac{2n\pi}{a}(ix-y)-\frac{1}{2}(b\sin\theta)^2\kappa h_a n^2]\right|^2 \quad (50)$$

with $h_a ab\sin\theta=\frac{2\pi}{\kappa}$ (51)

is the solution of Eq.(49) in whole space [33], where $C_0$ is a constant, being required to fit the condition $2\kappa h_a >> f$ ; $a$ and $\theta$ are two parameters that can be adjusted. The relation $h_a ab\sin\theta = 2\pi/\kappa$ show that there is one of fluxon through each space cell ( $ab\sin\theta$ ), suggesting that the magnetic field is at mixed state of type II superconductors. Figure 2 shows the contour plots of magnetic field $h$ for three set of different parameters, in which $\kappa=7$ , $h_a=6.9$ and (a) a = 0.4, $\theta=\pi/2$ ; (b) $a=b=2\cdot 3^{-1/4}(\kappa h_a/\pi)^{-1/2}$ = 0.3875708032, $\theta=\pi/3$; (c) a = 0.4, $\theta=\pi/4$, respectively. Here we set $C_0=1$ that fits condition $2\kappa h_a >> f$ . In Fig. 1, picture (b) corresponds to the minimum of relative error $<(f-\bar{f})^2>/\bar{f}^2=<f^2>/<f>^2-1$ , namely, the $f$ varies in the slowest and smoothest manner. It has ever been related to the minimum free energy under Ginzburg-Landau theory [32].

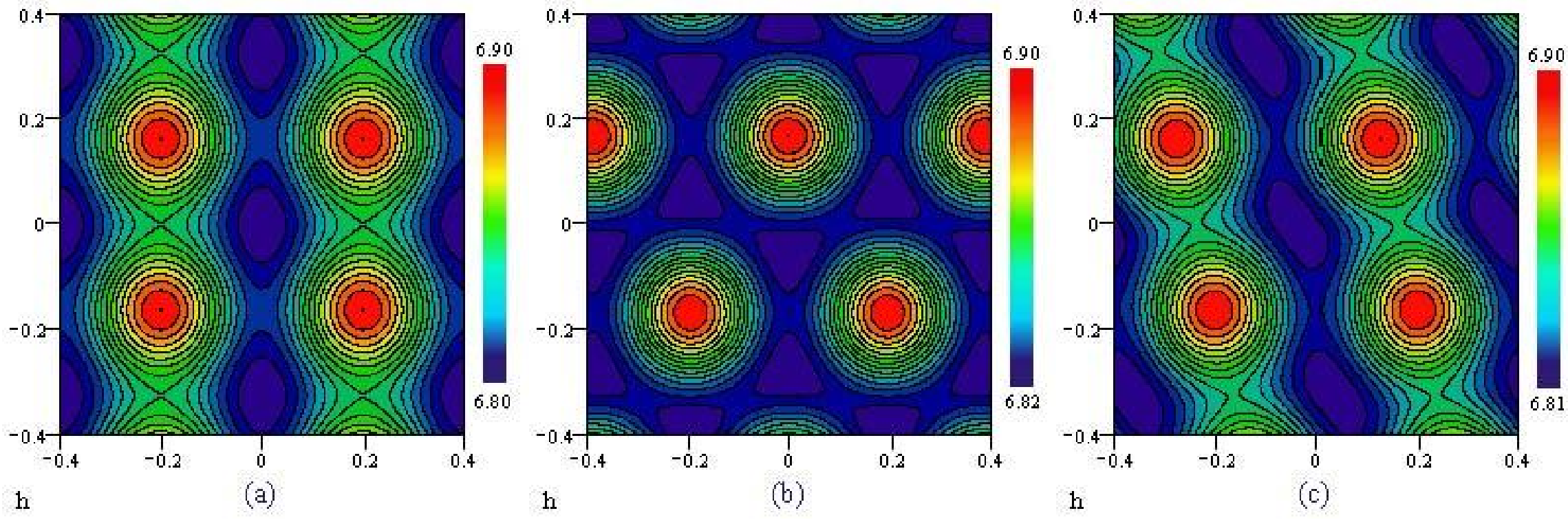

Fig. 2 The numerical results of magnetic field $h$ at mixed state of type II superconductors for three sets of different parameters. (a) a = 0.4, $\theta=\pi/2$; (b)a = 0.3875708032, $\theta=\pi/3$; (c) a = 0.4, $\theta=\pi/4$. $\kappa=7$ and $h_a=6.9$ for all.

## 4. Conclusion

In conclusion, we have proposed a set of modified electromagnetic equations in isotropic materials, from which the Ohm law is derived. And the dissipation of current is found to be with the form $\mathbf{J}\cdot\mathbf{E} = c^2\rho/(u\varepsilon_r\mu_r)$. This leads to a condition of free charge density $\rho = 0$ other than zero resistance for superconductivity. The condition $\rho = 0$ characterizes the superconducting states different from those described by London theory ( $\alpha = \text{constant}$ ). However, whether $\rho = 0$ or $\alpha = \text{constant}$, the Ohm law is always invalid. So the line lying between presence and absence of Ohm law is the boundary of regular- and super-conductive materials. It is also found that in a steady state, when $\mathbf{J} \neq 0$ but $\rho = 0$, it must be $\mathbf{E} = 0$ i.e., all electric fields involving Hall one will vanish, which suggests a superconductive pairing mechanism of holes and electrons. This solves the pending problem why vanishing of Hall-effect in some superconducting states.

**Acknowledgment**

The author thanks Professor Binglong Chen for helpful discussion on differential geometry.

# Appendix

We now begin to deduce the Eqs. (1)-(6) basing on the charge conservation law and differential geometry, with consulting Maxwell equations. Here we consider the isotropic materials, in which the electromagnetic field has no singularity, i.e., the space-time manifold related to the medium is a simply connected one. In this space-time manifold, the charge conservation law

$$\nabla\cdot\mathbf{J} + \frac{\partial\rho}{\partial t} = 0 \quad \text{(A1)}$$

can be expressed as

$$d\omega_3 = 0, \quad \text{(A2)}$$

where $d$ is the exterior differential operator [34, 35] and $\omega_3 = R_{123}dx^{123} + R_{012}dx^{012} + R_{013}dx^{013} + R_{023}dx^{023}$ is a 3-form; $dx^{123} = dx^1 \wedge dx^2 \wedge dx^3$ , $dx^{012} = dx^0 \wedge dx^1 \wedge dx^2$ , $dx^{013} = dx^0 \wedge dx^1 \wedge dx^3$ and $dx^{023} = dx^0 \wedge dx^2 \wedge dx^3$ ( $x^0 = ct$ , and in a rectangular coordinate system, $x^1 = x$ , $x^2 = y$ , $x^3 = z$ . $R_{123} = c\rho$ , $R_{012} = -J_z$ , $R_{013} = J_y$ , $R_{023} = -J_x$ ). Given Eq. (A2), the charge conservation law becomes

$$R_{123,0} - R_{012,3} + R_{013,2} - R_{023,1} = 0, \quad \text{(A3)}$$

where $R_{123\ ,0} = \partial R_{123}/\partial x^0$, $R_{012\ ,3} = \partial R_{012}/\partial x^3$, $R_{013\ ,2} = \partial R_{013}/\partial x^2$, $R_{023\ ,1} = \partial R_{023}/\partial x^1$ . From Eq.(A2) we know that $\omega_3$ is a closed form. According to Poincaré lemma [34] , 3-form $\omega_3$ should also be an exact form, which means that there exists a non-closed 2-form $\omega_2 = G_{01}dx^{01} + G_{02}dx^{02} + G_{03}dx^{03} + G_{12}dx^{12} + G_{13}dx^{13} + G_{23}dx^{23}$ making

$\omega_3 = d\omega_2$ (A4)

or

$R_{123} = G_{12,3} - G_{13,2} + G_{23,1}$ (A5)

$R_{012} = G_{01,2} - G_{02,1} + G_{12,0}$ (A6)

$R_{013} = G_{01,3} - G_{03,1} + G_{13,0}$ (A7)

$R_{023} = G_{02,3} - G_{03,2} + G_{23,0}$ (A8)

One can see that Eq. (A5) is just Maxwell's equation $\nabla \cdot \mathbf{D} = \rho$ and Eqs. (A6)-(A8) are equivalent to Maxwell's equation $\nabla \times \mathbf{H} = \mathbf{J} + \partial \mathbf{D} / \partial t$, where

$\mathbf{H} = (G_{01},\ G_{02},\ G_{03}),\ c\mathbf{D} = (G_{23},\ -G_{13},\ G_{12})$. (A9)

In fact, in the 4-dimension space-time manifold, besides $\omega_2$ it permits that there exists a closed 2-form $\tau_2 = F_{01}dx^{01} + F_{02}dx^{02} + F_{03}dx^{03} + F_{12}dx^{12} + F_{13}dx^{13} + F_{23}dx^{23}$ ( $d\tau_2 = 0$ ), which makes the following equation held:

$\omega_3 = d(\omega_2 + \tau_2)$ . (A10)

Of course we have

$F_{12,3} - F_{13,2} + F_{23,1} = 0$ (A11)

$F_{01,2} - F_{02,1} + F_{12,0} = 0$ (A12)

$F_{01,3} - F_{03,1} + F_{13,0} = 0$ (A13)

$F_{02,3} - F_{03,2} + F_{23,0} = 0$ (A14)

According to Poincaré lemma [34], 2-form $\tau_2$ should be an exact form, which means that there exists a non-closed 1-form $\omega_1 = A_0dx^0 + A_1dx^1 + A_2dx^2 + A_3dx^3$ that makes

$\tau_2 = d\omega_1$ . (A15)

It immediately leads to

$$\begin{array}{lll} F_{01} = A_{1,0} - A_{0,1}\,, & F_{02} = A_{2,0} - A_{0,2}\,, & F_{03} = A_{3,0} - A_{0,3} \\ F_{12} = A_{2,1} - A_{1,2}\,, & F_{13} = A_{3,1} - A_{1,3}\,, & F_{23} = A_{3,2} - A_{2,3} \end{array} \text{. (A16)}$$

In the 4-dimension space-time manifold, besides $\omega_1$ it permits that there exists a closed 1-form $\tau_1 = C_0dx^0 + C_1dx^1 + C_2dx^2 + C_3dx^3$ ( $d\tau_1 = 0$ ), which makes the following equation held:

$\tau_2 = d(\omega_1 + \tau_1)$ (A17)

And equation $d\tau_1 = 0$ reads

$$\begin{array}{lll} C_{1,0} - C_{0,1} = 0\,, & C_{2,0} - C_{0,2} = 0\,, & C_{3,0} - C_{0,3} = 0 \\ C_{2,1} - C_{1,2} = 0\,, & C_{3,1} - C_{1,3} = 0\,, & C_{3,2} - C_{2,3} = 0 \end{array} \text{. (A18)}$$

Again, according to Poincaré lemma [34], 1-form $\tau_1$ should be an exact form, which means that there exists a non-closed 0-form $\phi$ making

$\tau_1 = d\phi$ . (A19)

Namely,

$C_0 = \phi_{,0}$ , $C_1 = \phi_{,1}$ , $C_2 = \phi_{,2}$ , $C_3 = \phi_{,3}$ (A20)

Up to this point, we have not discussed the physics of $\tau_2, \omega_1, \tau_1$ and $\phi$ yet. To endow these forms with physical meaning, we now build the relations between $\tau_2$ and $\omega_2$ as well as that between $\omega_3$ and $(\omega_1 + \tau_1)$ with the aid of Hodge * operation [35]. Hodge * operation is an algebraic operation stock that maps an r-form onto a (n-r)-form (r<n) and depends on the metric of space-time manifold considered. In our case, n = 4, so the Hodge * operation can map a 2-form onto a 2-form as well as map a 3-form onto a 1-form. To perform the Hodge * operation we need to introduce a metric for our space-time manifold. For convenience and without losing the generality we consider the following metric for an isotropic medium:

$$g^{\lambda\nu} = \begin{cases} -\mu_r\varepsilon_r & \lambda = \nu = 0 \\ 1 & \lambda = \nu = 1,2,3 \\ 0 & others \end{cases} . \quad \text{(A21)}$$

With the metric we can map the non-closed 2-form $\omega_2$ onto a closed 2-form $\beta_2$ ( $d\beta_2 = 0$ ), i.e., $*\omega_2 = \beta_2$ . The $\beta_2$ would not be just equal to $\tau_2$ . But it is possible to introduce a parameter $f$ making

$*\omega_2 = \beta_2 = f\tau_2$ . (A22)

where $f$ depends on space coordinates and time. Under the metric described by Eq. (A21) (also see the note below), we know that

$$*\omega_2 = \sqrt{-g}(G_{23}dx^{01} - G_{13}dx^{02} + G_{12}dx^{03} + g^{00}G_{03}dx^{12} - g^{00}G_{02}dx^{13} + g^{00}G_{01}dx^{23}) , \quad \text{(A23)}$$

where $\sqrt{-g} = \sqrt{\mu_r\varepsilon_r}$ and $g^{00} = -\varepsilon_r\mu_r$ . Using Eqs. (A9) , (A22) and (A23) we obtain

$$\begin{aligned} \tau_2 &= F_{01}dx^{01} + F_{02}dx^{02} + F_{03}dx^{03} + F_{12}dx^{12} + F_{13}dx^{13} + F_{23}dx^{23} \\ &= \frac{\sqrt{-g}}{f}(G_{23}dx^{01} - G_{13}dx^{02} + G_{12}dx^{03} + g^{00}G_{03}dx^{12} - g^{00}G_{02}dx^{13} + g^{00}G_{01}dx^{23}) \\ &= \frac{\sqrt{\mu_r\varepsilon_r}}{f}(cD_1dx^{01} + cD_2dx^{02} + cD_3dx^{03} - \mu_r\varepsilon_rH_3dx^{12} + \mu_r\varepsilon_rH_2dx^{13} - \mu_r\varepsilon_rH_1dx^{23}) \end{aligned} \quad \text{. (A24)}$$

Choosing parameter $f = -\varepsilon_r\sqrt{\mu_r\varepsilon_r} / \mu_0$ , we get from Eq. (A24) that

$$\begin{aligned} &F_{01} = \mu_0\varepsilon_0cE_1 , \; F_{02} = \mu_0\varepsilon_0cE_2 , \; F_{03} = \mu_0\varepsilon_0cE_3 \\ &F_{12} = -B_3 , \quad F_{13} = B_2 , \quad F_{23} = -B_1 \end{aligned} . \quad \text{(A25)}$$

Therefore, Eq. (A11) becomes as

$\nabla \cdot \mathbf{B} = 0$ , (A26)

and Eqs. (A12) - (A14) as

$\nabla\times\mathbf{E}=-\partial\mathbf{B}/\partial t$ . (A27)

Equations (A26) and (A27) are just two Maxwell's equations. This confirms the existence of relation (A22), and shows the validity of using differential geometry method here. One can find that the equations involved in (A16) are equivalent to $\mathbf{B}=\nabla\times\mathbf{A}$ and $\mathbf{E}=-\partial\mathbf{A}/\partial t-\nabla\varphi$ , where $\varphi=-cA_0$ .

Now that the Hodge * operation mapping the 2-form $\omega_2$ onto a 2-form $\beta_2$ can induce correct Eqs (A26) and (A27), we wish that the operation mapping the 3-form $\omega_3$ onto a 1-form $\beta_1$ would induce other useful relations. To this end, we introduce another parameter *k* depending on space coordinates and time, which makes

$$*\omega_3=\beta_1=k(\omega_1+\tau_1)\quad .\ (\text{A28})$$

Under the metric mentioned above (also see the note below),

$$*\omega_3=\sqrt{-g}\,(-R_{123}dx^0+g^{00}R_{023}dx^1-g^{00}R_{013}dx^2+g^{00}R_{012}dx^3)\ .\ (\text{A29})$$

On the other hand (see the expressions of $\omega_1$ and $\tau_1$ above),

$$\omega_1+\tau_1=(A_0+C_0)dx^0+(A_1+C_1)dx^1+(A_2+C_2)dx^2+(A_3+C_3)dx^3\ .\ (\text{A30})$$

Therefore,

$$k(A_0+C_0)=-\sqrt{\mu_r\varepsilon_r}R_{123}=-\sqrt{\mu_r\varepsilon_r}c\rho,\ \ k(A_1+C_1)=-\mu_r\varepsilon_r\sqrt{\mu_r\varepsilon_r}R_{023}=\mu_r\varepsilon_r\sqrt{\mu_r\varepsilon_r}J_x$$

$$k(A_2+C_2)=\mu_r\varepsilon_r\sqrt{\mu_r\varepsilon_r}R_{013}=\mu_r\varepsilon_r\sqrt{\mu_r\varepsilon_r}J_y,\ \ k(A_3+C_3)=-\mu_r\varepsilon_r\sqrt{\mu_r\varepsilon_r}R_{012}=\mu_r\varepsilon_r\sqrt{\mu_r\varepsilon_r}J_z$$

(A31)

By using Eq(A20), these equations can be rewritten as

$$\rho=\frac{k}{c^2\sqrt{\mu_r\varepsilon_r}}(\varphi-\frac{\partial\phi}{\partial t})\quad ,\ (\text{A32})$$

$$\mathbf{J}=\frac{k}{\mu_r\varepsilon_r\sqrt{\mu_r\varepsilon_r}}(\mathbf{A}+\nabla\phi)\quad .\ (\text{A33})$$

One sees that the Hodge * operation has led us to get two equations $\nabla\cdot\mathbf{B}=0$ and $\nabla\times\mathbf{E}=-\partial\mathbf{B}/\partial t$ , in which the fields $\mathbf{E}$ and $\mathbf{B}$ satisfy Maxwell's equations $\nabla\cdot\mathbf{D}=\rho$ and $\nabla\times\mathbf{H}=\mathbf{J}+\partial\mathbf{D}/\partial t$ , also, get Eqs.(A32) and (A33), building the relations between $\rho$ and $\varphi$ , $\mathbf{J}$ and $\mathbf{A}$ .

In fact, the above four equations (A4), (A17), (A22) and (A28) will lead to this relation

$$d[\frac{1}{k}(*d\omega_2)]=\frac{1}{f}*\omega_2\quad ,\ (\text{A34})$$

which involves six differential equations that can, in principle, determine six parameters *f*, *k* and $g^{\mu\mu}$ $(\mu=0,1,2,3)$ if the 2-form $\omega_2$ (or the electromagnetic field) has been given.

**Note:**

$$
\begin{aligned}
*\omega_2 &= *(G_{01}dx^{01} + G_{02}dx^{02} + G_{03}dx^{03} + G_{12}dx^{12} + G_{13}dx^{13} + G_{23}dx^{23}) \\
&= G_{01}\;{}^{01}_{23}\,dx^{23} + G_{02}\;{}^{02}_{13}\,dx^{13} + G_{03}\;{}^{03}_{12}\,dx^{12} + G_{12}\;{}^{12}_{03}\,dx^{03} + G_{13}\;{}^{13}_{02}\,dx^{02} + G_{23}\;{}^{23}_{01}\,dx^{01} \\
&= G_{01}g^{0\mu}g^{1\nu}\;{}_{\mu\nu 23}\,dx^{23} + G_{02}g^{0\mu}g^{2\nu}\;{}_{\mu\nu 13}\,dx^{13} + G_{03}g^{0\mu}g^{3\nu}\;{}_{\mu\nu 12}\,dx^{12} \\
&\quad + G_{12}g^{1\mu}g^{2\nu}\in_{\mu\nu 03}dx^{03} + G_{13}g^{1\mu}g^{3\nu}\;{}_{\mu\nu 02}\,dx^{02} + G_{23}g^{2\mu}g^{3\nu}\;{}_{\mu\nu 01}\,dx^{01} \\
&= G_{01}g^{00}g^{11}\;{}_{0123}\,dx^{23} + G_{02}g^{00}g^{22}\;{}_{0213}\,dx^{13} + G_{03}g^{00}g^{33}\;{}_{0312}\,dx^{12} \\
&\quad + G_{12}g^{11}g^{22}\;{}_{1203}\,dx^{03} + G_{13}g^{11}g^{33}\;{}_{1302}\,dx^{02} + G_{23}g^{22}g^{33}\;{}_{2301}\,dx^{01} \\
&= \sqrt{-g}(G_{01}g^{00}dx^{23} - G_{02}g^{00}dx^{13} + G_{03}g^{00}dx^{12} + G_{12}dx^{03} - G_{13}dx^{02} + G_{23}dx^{01})
\end{aligned}
$$

$$
\begin{aligned}
*\omega_3 &= *(R_{123}dx^{123} + R_{012}dx^{012} + R_{013}dx^{013} + R_{023}dx^{023}) \\
&= R_{123}\in_0^{123}dx^0 + R_{012}\in_3^{012}dx^3 + R_{013}\in_2^{013}dx^2 + R_{023}\in_1^{023}dx^1 \\
&= R_{123}g^{1\mu}g^{2\nu}g^{3\lambda}\in_{\mu\nu\lambda 0}dx^0 + R_{012}g^{0\mu}g^{1\nu}g^{2\lambda}\in_{\mu\nu\lambda 3}dx^3 \\
&\quad + R_{013}g^{0\mu}g^{1\nu}g^{3\lambda}\in_{\mu\nu\lambda 2}dx^2 + R_{023}g^{0\mu}g^{2\nu}g^{3\lambda}\in_{\mu\nu\lambda 1}dx^1 \\
&= R_{123}g^{11}g^{22}g^{33}\in_{1230}dx^0 + R_{012}g^{00}g^{11}g^{22}\in_{0123}dx^3 \\
&\quad + R_{013}g^{00}g^{11}g^{33}\in_{0132}dx^2 + R_{023}g^{00}g^{22}g^{33}\in_{0231}dx^1 \\
&= \sqrt{-g}(-R_{123}dx^0 + R_{012}g^{00}dx^3 - R_{013}g^{00}dx^2 + R_{023}g^{00}dx^1)
\end{aligned}
$$